\documentclass{webofc}
\usepackage[varg]{txfonts}   
\woctitle{QUARKS-2016, 19th International Seminar on High Energy Physics, Pushkin, Russia, 29 May - 4 June, 2016}

\newcommand{\rar}{\rightarrow}
\newcommand{\lrar}{\leftrightarrow}

\def\mc{\mathcal}
\def\be{\begin{eqnarray}}
\def\ee{\end{eqnarray}}

\def\D{\mc D}
\def\-g{\sqrt{-g}}
\begin{document}
\title{Kinetics of spontaneous baryogenesis in non-stationary background}
%
%

\author{\firstname{Elena} \lastname{Arbuzova}\inst{1,2}\fnsep\thanks{\email{arbuzova@uni-dubna.ru
    }} \and
        \firstname{Alexander} \lastname{Dolgov}\inst{1,3,4}\fnsep\thanks{\email{dolgov@fe.infn.it
             }} \and
        \firstname{Victor} \lastname{Novikov}\inst{3}\fnsep\thanks{\email{novikov@itep.ru
             }}
      }

\institute{Novosibirsk State University, Novosibirsk, 630090, Russia
\and
           Department of Higher Mathematics, Dubna State University, 141980 Dubna, Russia
\and
           ITEP, Bol. Cheremushkinsaya ul., 25, 117259 Moscow, Russia
\and    
Dipartimento di Fisica e Scienze della Terra, Universit\`a degli Studi di Ferrara\\
Polo Scientifico e Tecnologico - Edificio C, Via Saragat 1, 44122 Ferrara, Italy           
          }

\abstract{%
 Generation of the cosmological  baryon asymmetry in frameworks of  spontaneous baryogenesis is studied in detail. It 
is shown that the relation between baryonic chemical potential and the time derivative of the (pseudo)Goldstone field
essentially depends upon the representation chosen for the fermionic fields with non-zero baryonic number (quarks).
Kinetic equation is modified and numerically solved in equilibrium
for the case of time dependent external background or finite integration time to be applicable to the case when energy 
conservation law is formally violated. 
}
\maketitle
\section{Introduction}
\label{intro}

One of the popular scenarios of baryogenesis is the spontaneous baryogenesis (SBG) proposed in 
papers~\cite{spont-BG-1,spont-BG-2,spont-BG-3}, for reviews see e.g. Refs.~\cite{BG-rev,AD-30}. 
It is assumed that in the unbroken phase the theory is
invariant with respect to the global $U(1)$-symmetry, which ensures conservation of baryonic number.
This symmetry is spontaneously broken and in the broken phase
the Lagrangian density  acquires the term
\begin{equation}
{\cal L}_{SB} =  (\partial_{\mu} \theta) J^{\mu}_B\, ,
 \label{L-SB}
 \end{equation}
 where $\theta$ is the Goldstone field and $J^{\mu}_B$ is the baryonic current. Due to the 
 spontaneous symmetry breaking (SSB) this current 
 is not conserved. The next step  is the statement that the Hamiltonian density corresponding to ${\cal L}_{SB}$ is 
 simply the Lagrangian density taken with the opposite sign:
 \begin{equation} 
{\cal H}_{SB} = - {\cal L}_{SB} = - (\partial_{\mu} \theta) J^{\mu}_B\, .
\label{H-SB}  
\end{equation}   
For the spatially homogeneous field $\theta = \theta (t)$ this Hamiltonian is reduced to ${\cal H}_{SB} =  - \dot \theta\, n_B$, where $n_B\equiv J^4_B$ 
is the baryonic number density, so it is tempting to identify $\dot \theta$ with the chemical potential, $\mu$, of the corresponding system. 
If this is the case, then in thermal equilibrium the baryon asymmetry would evolve to:
\begin{equation} 
n_B =\frac{g_S B_Q}{6} \left({\mu\, T^2} + \frac{\mu^3}{ \pi^2}\right) \rightarrow
\frac{g_S B_Q}{6} \left({\dot \theta\, T^2} + \frac{\dot \theta ^3}{ \pi^2}\right)\,,
\label{n-B-of-theta}
\end{equation}
where  $T$ is the cosmological plasma temperature, $g_S$ and $B_Q$ are respectively the number of the spin states 
and the baryonic number of quarks, which are supposed to be the bearers of the baryonic number. 

It is interesting that for successful SBG two of the three Sakharov's conditions for the generation of the cosmological 
baryon asymmetry, namely, breaking of thermal equilibrium and a violation of C and CP symmetries are 
unnecessary.  This scenario is analogous the baryogenesis in absence of CPT invariance, if the masses of particles and antiparticles are
different. In the latter case the generation of the cosmological baryon asymmetry can also proceed in thermal equilibrium~\cite{ad-zeld-cpt,ad-cpt}.

In this work the classical version of spontaneous baryogenesis  is studied. The talk is organized as follows.  In Section~\ref {s-ssb} the general features of the spontaneous breaking of baryonic $U(1)$-symmetry are described, and the (pseudo)Goldstone  mode, its equation of motion, and baryonic chemical potential are introduced. 
Next, in Sec.~\ref{s-kin-eq-canon} the standard kinetic 
equation in stationary background is presented. 
In Sec.~\ref{s-kin-eq} we derive kinetic equation
in time dependent external field and/or for the case when energy is not conserved because of finite limits of integration over time. Several
examples, when such kinetic equation is relevant, are presented in Sec.~\ref{s-examples}. Lastly in Sec.~\ref{s-conclude} we conclude.

\section{Spontaneous symmetry breaking and goldstone mode \label{s-ssb}}

Let us consider the theory of complex scalar field $\Phi$ interacting with "quarks", $Q$, and "leptons", $L$, 
with the Lagrangian:
\be
{\cal L }(\Phi) =  g^{\mu\nu} \partial_\mu \Phi^*
\partial_\nu \Phi - V(\Phi^* \Phi) + \bar Q (i \gamma^\mu \partial_\mu - m_Q)\,Q 
+  \bar L ( i \gamma^\mu \partial_\mu - m_L) L + {\cal L}_{int}(\Phi, Q, L)\, ,
\label{S-Phi}
\ee
where ${\cal L}_{int} $ describes the interaction between $\Phi $ and fermionic fields. In the toy model studied
below we take it in the form:
\be
{\cal L}_{int} =  \frac{\sqrt 2}{m_X^2} \frac{\Phi}{f}\, (\bar L \gamma_{\mu} Q )(\bar Q^c  \gamma_{\mu} Q) +
h.c. \, , 
\label{L-int}
\ee
where $Q^c$ is charged conjugated quark spinor, 
$m_X$ is a parameter with dimension of mass, and $f$ is related to the vacuum expectation value of $\Phi $ defined below in Eq.~(\ref{V-of-Phi}).
Such an interaction can appear e.g. in $SU(5)$ Grand Unified Theory.  For simplicity, in our toy model  we do not take into account the quark colors. 

B-non conserving  interaction may have many different forms. The one presented above describes transition of three quark-type fermions into 
(anti)lepton. There may be transformation of two or three quarks into equal number of antiquarks. Such interaction describes neutron-antineutron
oscillations. 
There even can be a "quark" transition into three "leptons". Depending on the interaction type
the relation between $\dot\theta$ and the effective chemical potential would have different forms.

Note that $Q$ and $L$ can be any fermions, not necessarily
quarks and leptons of the standard model.  For example, they can
be new heavy fermions. They may possess similar or the same quantum numbers as the quarks and leptons of the standard model 
and may couple to the ordinary quarks and leptons.  In section~\ref{s-kin-eq} we consider another model to  study kinetics of the baryon
asymmetry generation which allows for the transformation $3L \lrar Q$ or
$2 Q \lrar 2\bar Q$.  They are surely not permitted for the standard quarks. However, the process $3 q \lrar 3\bar q$ is permitted and kinetics
of this process is essentially the same. We denote by $q$ the fermionic field with the same quantum number as the usual quark. 

The theory (\ref{S-Phi}) considered in this section is invariant under the following $U(1)$-transformations:
\be
\Phi \rightarrow e^{i \alpha} \Phi, ~~~~Q \rightarrow e^{- i \alpha/3} Q,
{}~~~~L \rightarrow L \, .
\label{phase}
\ee
In the unbroken symmetry phase this invariance leads to the conservation of the total baryonic number which includes 
the baryonic number of $\Phi $, taken to be unity,
and that of quarks, equal to $1/3$. In realistic model the interaction of left- and right-handed fermions may be different but 
we neglect this possible difference in what follows.

We assume that the global $U(1)$-symmetry is spontaneously broken at the
energy scale $f$ in the usual way, e.g. via the potential of the form
\be
V(|\Phi|) = \lambda \left(\Phi^* \Phi - f^2/2 \right)^2 \, .
\label{V-of-Phi}
\ee
The resulting scalar field vacuum expectation value is $\langle \Phi \rangle = f e^{i\phi_0/f}/\sqrt{2}$ with a constant phase $\phi_0$.

Below the scale $f$ we can neglect the heavy radial mode of
$\Phi$ with the mass $m_{radial} = \lambda^{1/2} f$, since being very
massive it is frozen out, but this simplification is not necessary and is not essential for the baryogenesis.
The remaining light degree of freedom is the variable field $\phi$, which is the Goldstone boson of the
spontaneously broken $U(1)$. Up to a constant factor the field $\phi$  
is the angle around the bottom of the Mexican hat potential described by Eq.~(\ref{V-of-Phi}). 
Correspondingly we introduce the dimensionless angular field $\theta \equiv \phi/f$:
\be
 \Phi  =f e^{i\phi / f} \sqrt{2}=f e^{i \theta} /\sqrt{2}\, .
 \label{theta}
 \ee

As a result the following effective Lagrangian for $\theta $ is obtained:
\be \nonumber
{\cal L}_1 (\theta)
= {f^2\over 2} \partial_\mu \theta \, \partial^{\mu} \theta + \bar Q_1
(i \gamma^{\mu} \partial_\mu - m_Q) Q_1 +  \bar L (
i \gamma^{\mu} \partial_\mu - m_L) L + \\ 
 \left(\frac{e^{i \theta }}{ m_X^2} \, (\bar L \gamma_{\mu} Q_1 )(\bar Q_1^c  \gamma_{\mu} Q_1) +
  h.c.\right) - 
 U(\theta)\, .
 \label{L-of-theta-1}
 \ee
Here we introduced "by hand"  potential  $ U(\theta) $, which may appear due to an  explicit symmetry 
breaking and can lead, in particular, to a nonzero mass of $\theta$. We use the notation $Q_1$ for the quark field to 
distinguish it from the phase rotated field $Q_2$ introduced below in Eq.~ (\ref{L-of-theta-2}). 
In a realistic model the quark fields should be (anti)symmetrized 
 with respect to color indices, omitted here for simplicity. 
 
 If $U(\theta ) = 0$, the theory still remains invariant under the global transformations (i.e. with $\alpha = const$):
 \be
Q \rightarrow e^{ - i \alpha/3} Q, ~~~~L \rightarrow L,
{}~~~~\theta \rightarrow \theta + \alpha  \, .
\label{global-trans}
\ee
If we only rotate the quark field as above  
 but with coordinate dependent $\alpha =  \theta (t, \bf x)$, introducing the new field $Q_1 = e^{- i \theta /3} Q_2$, then the    
 Lagrangian~(\ref{L-of-theta-1}) is transformed into:
\be
{\cal L}_2 (\theta)
= {f^2\over 2} \partial_\mu \theta \partial^{\mu} \theta + \bar Q_2
(i \gamma^{\mu} \partial_\mu - m_Q) Q_2 +  \bar L (
i \gamma^{\mu} \partial_\mu - m_L) L + \nonumber \\
 \left(\frac{1}{ m_X^2} \, (\bar Q_2 \gamma_{\mu} L )(\bar Q_2  \gamma_{\mu} Q_2^c) +
  h.c.\right) + (\partial_\mu \theta) J^\mu - U(\theta) \, , 
  \label{L-of-theta-2}
 \ee
where the quark baryonic current is $J_\mu =(1/3) \bar Q \gamma_\mu Q$. Note that the current has the same form in terms of $Q_1$ and $Q_2$.

The equation of motion for the quark field $Q_1$ obtained from  Lagrangian (\ref{L-of-theta-1}) has the form:
\be
(i \gamma^{\mu} \partial_\mu - m_Q) Q_1 +  \frac{e^{-i \theta }}{ m_X^2} \left[  \gamma_{\mu} L (\bar Q_1  \gamma_{\mu} Q_1^c) +
2 \gamma_{\mu} Q_1^c (\bar Q_1 \gamma_{\mu} L )  \right]  =0\,.
  \label{dirac-1}
 \ee
Analogously the equation of motion for the phase rotated field $Q_2$ derived from Lagrangian (\ref{L-of-theta-2})  is: 
\be
\left(i \gamma^{\mu} \partial_\mu - m_Q + \frac{1}{3}\,\gamma ^\mu \partial_\mu \theta \right) Q_2 
+  \frac{1}{ m_X^2} \left[  \gamma_{\mu} L (\bar Q_2  \gamma_{\mu} Q_2^c) +
2 \gamma_{\mu} Q_2^c (\bar Q_2 \gamma_{\mu} L )  \right]  =0\,.
  \label{dirac-2} 
 \ee
Equations for $\theta$-field derived from these two Lagrangians in flat space-time have respectively the forms:
\be 
f^2 (\partial _t^2 - \Delta ) \theta + U'(\theta) 
+ \left[ \frac{i \,e^{- i \theta }}{ m_X^2} \, (\bar Q_1 \gamma_{\mu} L )(\bar Q_1  \gamma_{\mu} Q_1^c) +
  h.c.\right] = 0 
  \label{d2-theta-1}
  \ee 
  and
  \be 
f^2 (\partial _t^2 - \Delta ) \theta + U'(\theta)  +  \partial _\mu J^\mu_B = 0\, ,
   \label{d2-theta-2}
  \ee 
  where $U'(\theta) = dU/d\theta $. 
  
 Using either the equation of motion (\ref{dirac-1}) or (\ref{dirac-2}) we can check that the baryonic
 current is not conserved. Indeed, its divergence is:
 \be
 \partial_\mu J^\mu_B = \frac{i \,e^{-i\theta}}{m_X^2} (\bar Q_1 \gamma_\mu Q_1^c) (\bar Q_1 \gamma^\mu L) 
 + h.c.
 \label{dmu-Jmu}
 \ee 
 (and similarly for $Q_2$ but without the factor $\exp(-i\theta)$).
 So the equations of motion for $\theta$ in both cases (\ref{d2-theta-1}) and (\ref{d2-theta-2}) coincide, as expected.

In the spatially homogeneous case, when $\partial_\mu J^\mu_B = \dot n_B $ and
$\theta = \theta (t)$, and if $U(\theta) = 0$,  equation (\ref{d2-theta-2}) can be easily integrated giving:
\be
f^2 \left[\dot \theta (t) - \dot \theta (t_{in})\right] =  -n_B (t) +  n_B (t_{in})  \,.
\label{nB-of-t}
\ee
It is usually assumed that the initial baryon asymmetry vanishes, $n_B(t_{in}) = 0$. 

The evolution of $n_B (t)$ is governed by the kinetic equation discussed in Sec.~\ref{s-kin-eq-canon}, which allows to express $n_B$ through $\theta (t)$
and thus to obtain the closed systems of, generally speaking, integro-differential equations. In thermal equilibrium the relation between
$\dot\theta$ and $ n_B $ may become an algebraic one, but this is true only in the case when the integration over time is sufficiently long and
 if $\dot\theta$ is constant or slowly varying  function of  time.

In cosmological Friedmann-Robertson-Walker (FRW) background and space-independent $\theta (t)$ equation (\ref{d2-theta-2}) is transformed to:
  \be
  f^2 (\partial _t + 3 H  ) \dot\theta + U'(\theta)   =  - (\partial _t + 3 H ) n_B.
  \label{d2-theta-FRW}
  \ee 
We do not include the curvature effects into the Dirac equations because this is not necessary for what follows. Still we are using expression
for the current divergence in the form $\D_\mu J^\mu =\dot n_B + 3 Hn_B$, but not just $\dot n_B$.

If particles (fermions)  are in thermal equilibrium with respect to baryo-conserving interactions, then their phase space distribution 
has the form:
\be
f_{eq} =\left[ 1 +  \exp (E/T -\xi_B) \right]^{-1},
\label{f-eq-ferm}
\ee
where dimensionless chemical potential $\xi_B$ has equal magnitude but opposite signs for particles and antiparticles.
The baryonic number density, for small $\xi_B$, is usually given by the expression
\be 
n_B = g_S  B_Q \xi_B T^3 /6 
\label{nB-of-xi}
\ee
(compare to Eq.~(\ref{n-B-of-theta})). However, the relation (\ref{nB-of-xi}) between the baryonic number density and chemical potential 
is true only for the normal relation between
the energy and three-momentum, $E=\sqrt{ p^2 + m^2}$. This is not the case if the dispersion relation has the form
\be 
E  = \sqrt{p^2 +m^2} \pm \dot\theta/3,
\label{E-split}
\ee
derived from the equation of motion (\ref{dirac-2}), where the signs $\pm$ refer to particles or antiparticles
respectively, as we see a little below. We should note that the above dispersion relation is derived under assumption of constant or slow varying
$\dot\theta$. Otherwise the Fourier transformed Dirac equation cannot be reduced to the algebraic one.

If the baryon number is conserved, $n_B$ remains constant in comoving volume and it means in turn that $\xi_B = const$ for massless
particles. If and when non-conservation of baryons is switched on, $\xi_B$ evolves according to kinetic equation. Complete thermal equilibrium
in the standard theory demands $n_B \rar 0$, but a deviation from thermal equilibrium of B-nonconserving interaction leads to generation of 
non-zero $\xi_B$ and correspondingly to non-zero $n_B$. As we will see in Sec.~\ref{s-kin-eq},  SBG allows for generation of
nonzero baryonic number in complete thermal equilibrium.

In terms of $\xi_B$ and the new function $\eta = \dot \theta / T^3$
equation (\ref{d2-theta-FRW})  takes the same form as eq. (\ref{nB-of-t}):
\be
f^2  \left[ \eta (t) - \eta (t_{in}) \right] = - \frac{ g_S B_Q}{6} \left[  \xi_B (t) - \xi_B (t_{in}) \right]
\label{xi-B-of-eta}
\ee
and thus
\be
f^2  \left[ \frac{\dot\theta (t)}{T^3(t) }- \frac{\dot\theta (t_{in})}{T_{in}^3} \right] = - \frac{ g_S B_Q}{6} \left[  \xi_B (t) - \xi_B (t_{in}) \right] .
\label{dot-theta-of-xi}
\ee
As we have already mentioned, $n_B (t_{in} ) = 0$, so according to Eq.~(\ref{nB-of-xi}) we should also take 
 $\xi_B (t_{in}) = 0$. However this initial condition for chemical potential is not true in the 
theory with the Lagrangian (\ref{L-of-theta-2}) and the Dirac equation (\ref{dirac-2}) for the quark field,
though the condition $n_B (t_{in} ) = 0$ is supposed to be always valid. Indeed,
the B-nonconserving interaction now conserves energy and thus this process does not split the energies of quarks and antiquarks.
However, these energies are split from the very beginning due to relation (\ref{E-split}). Correspondingly using eqs. (\ref{f-eq-ferm}) 
and (\ref{E-split}) we find in the massless case:
\be 
n_B = \int \frac{d^3 p}{(2\pi)^3} (f_B - f_{\bar B} ) =
\frac{ g_S  B_Q}{6} \left(\xi_B - \frac{\dot\theta}{3T}\right)\,T^3 .
\label{nB-of-xi-2}
\ee
If initially $n_B = 0$, then $\xi_B (t_{in}) = \dot\theta_{in}/(3T_{in})$. In the case of conserved baryonic number, $n_B$ remains zero and thus in
equilibrium the relation $\xi_B = \dot\theta /(3T )$ must be true at any time. When the B-nonconserving interaction  is on, the
chemical potential would evolve and might evolve even down to zero, leading to generation of non-zero baryonic density,
as is discussed in Sec.~\ref{s-kin-eq}.  

In the pseudogoldstone case, when  $U(\theta) \neq 0$, equations of motion (\ref{d2-theta-2}) or (\ref{d2-theta-FRW}) 
cannot be so easily integrated, but in thermal equilibrium the system 
of equations containing $\theta (t)$ and $\xi_B(t)$ can be reduced to ordinary differential equations which are easily solved numerically.
Out of equilibrium one has to solve much more complicated system of the ordinary differential equation of motion for $\theta(t)$
and the integro-differential kinetic equation. It is discussed below in Sec.~\ref{s-kin-eq-canon}.

\section{Kinetic equation for time independent amplitude \label{s-kin-eq-canon}}

The temporal evolution of the distribution function 
of i-th type particle, $f_i (t,p)$, in an arbitrary process
${ i+Y \lrar Z}$ in the FRW background, is governed by the kinetic equation:
\be
\frac{df_i}{dt} = (\partial_t - H\,p_i \partial_{p_i}) f_i = I_i^{coll}, 
\label{kin-eq-1}
\ee
with the collision integral equal to:
\be 
&&I^{coll}_i={(2\pi)^4 \over 2E_i}  \sum_{Z,Y}  \int  \,d\nu_Z  \,d\nu_Y
\delta^4 (p_i +p_Y -p_Z)
\nonumber\\
&& \left[ |A(Z\rightarrow  i+Y)|^2
\prod_Z f \prod_{i+Y} (1\pm f) - 
 |A(i+Y\rightarrow  Z)|^2
 f_i  \prod_Y  f\prod_Z  (1\pm  f)\right],
\label{I-coll-1}
\ee
where $ A( a \rar b)$ is the amplitude of the transition from state $a$ to state $b$,
${ Y}$ and ${ Z}$ are  arbitrary, generally  multi-particle  states,
$\left( \prod_Y f \right)$ is  the  product  of  the phase space densities  of  particles forming the state $Y$, and
\be
d\nu_Y = \prod_Y {\overline {dp}} \equiv \prod_Y {d^3p\over 2E\,(2\pi )^3}.
\label{dnuy-1}
\ee
The signs '+' or '$-$' in ${ \prod (1\pm f)}$ are chosen for  bosons  and fermions respectively. We neglect the effects of the space-time
curvature  in the collision integral  which is generally a good approximation. 

In the lowest order of perturbation theory the amplitude of transition from an initial state $| in \rangle$ to a final state $| fin \rangle$ is
given by the integral of the matrix element of Lagrangian density between these states, integrated over 4-dimensional space $d^4 x$. 
The quantum field operators are expanded in terms of creation-annihilation operators with a plane wave coefficients: $\sim \exp (-i Et + i {\bf px} ) $.  

When the amplitude of the process is time-independent, then the integration of the product of the exponents in infinite integration limits 
leads to the energy-momentum conservation factors: 
\be
\int dt d^3 x \,e^{-i ( E_{in} - E_{fin} ) t + i (\bf P_{in} - \bf P_{fin} ) \bf x } = (2\pi)^4 \delta (E_{in} - E_{fin})\, \delta((\bf P_{in} - \bf P_{fin} ),
\label{int-over-d4x}
\ee
where $E_{in} $, $E_{fin}$,    $\bf P_{in}$, and $\bf P_{in}$ are the total energies and 3-momenta of the initial and final states respectively.
The amplitude squared contains delta-function of zero which is interpreted as the total time duration, $t_{max}$, of the process and as the
total space volume, $V$. The probability of the process given by the collision integral is normalized per unit time and volume, so it must be 
divided by $V$ and $t_{max}$.

We are interested in the evolution of the baryon number density, which is the time component of the baryonic current
$J^\mu$: $n_B \equiv J^4$. Due to the quark-lepton transitions the current is non-conserved and its divergence is
given by Eq.~(\ref{dmu-Jmu}). The similar expression is evidently true in terms of $Q_2$ but without the factor $\exp (- i \theta)$.
Let us first consider  the latter case, when
the interaction is described by the Lagrangian (\ref{L-of-theta-2}), which contains the product of three "quark" and one "lepton"
operators, and take as an example the process  $q_1 + q_2 \lrar \bar q + l $. 

Since the interaction in this representation does not depend on time, the energy is conserved and the collision integral has the 
usual form with conserved four-momentum. Quarks are supposed to be in kinetic equilibrium but probably not in equilibrium with 
respect to  B-nonconserving interactions, so their distribution function has the  form:
\be
f_Q = \exp \left( -\frac{E}{T} + \xi_B \right) \,\,\, {\rm and} \,\,\, f_{\bar Q} = \exp \left( -\frac{E}{T} - \xi_B \right) 
\label{f-B-equil}.
\ee
Here  and in what follows the Boltzmann statistics is used. 
Since the dispersion relation for quarks and antiquarks ~(\ref{E-split})
depends upon $\dot\theta$, the baryon asymmetry in this case is given by eq.~(\ref{nB-of-xi-2}) and  the kinetic equation takes the form:
\be
\frac{g_S  B_Q}{6}\, \frac{ d}{dt} \left( \xi_B - \frac{\dot\theta}{3T}\right) = - c_1\Gamma \xi_B,
\label{dot-n-B-xi}
\ee
where $c_1$ is a numerical factor of order unity and $\Gamma$ is the rate of baryo-nonconserving reactions.
If the amplitude of these reactions has the form presented in Eq.~(\ref{dirac-2}), then $\Gamma \sim T^5/m_X^4$.

For constant or slow varying temperature the equilibrium solution to this equation is $\xi_B = 0$ and the baryon number
density is proportional to $n_B \sim \dot\theta T^2 $, with $\dot\theta$ evolving according Eq.~(\ref{nB-of-t}) with $n_B$ expressed
through $\dot \theta$.

Let us check now what happens if the dependence on $\theta$ is moved from the quark dispersion relation to the B-nonconserving
interaction term (\ref{d2-theta-1}).
The expression for the collision integral (\ref{I-coll-1}) is valid only in absence of external field depending on coordinates. In our case, when 
quarks "live" in the $\theta (t)$-field, the collision integral should be modified in the following way. 
Now we have an additional factor under the integral (\ref{int-over-d4x}), namely, $\exp [ \pm i\theta (t)]$. In general case this integral cannot be 
taken analytically, but if we can approximate $\theta (t) $ as $\theta(t) \approx \dot \theta t$ with a constant or slowly varying 
$\dot \theta$, the integral is simply taken giving e.g. for the process of two quark transformation into antiquark and lepton,
$q_1+q_2 \lrar \bar q + l$, the energy balance condition imposed by
$\delta (E_{q_1} + E_{q_2} - E_{\bar q} - E_l - \dot\theta)$.
In other words the energy is non-conserved due to the action of the external field $\theta (t)$. The approximation of linear evolution of 
$\theta $ with time can be valid if the reactions are fast in comparison with the rate of the $\theta$-evolution. 

Returning to our case we can see that the collision integral
integrated over the  three-momentum of the particle under scrutiny
(i.e. particle $i$ in eq.~(\ref{I-coll-1}) )
 e.g. for process the $q_1+q_2 \rar  l + \bar q $ turns into:
\be  \nonumber
&&\dot n_B +3H n_B \sim \\
&&\int d\tau_{l \bar q} d\tau_{q_1 q_2} |A|^2 \delta (E_{q_1} + E_{q_2} - E_l - E_{\bar q} - \dot \theta) 
\delta( {\bf P}_{in} - {\bf P}_{fin} ) e^{- E_{in}/T} 
\left( e^{\xi_L - \xi_B + \dot \theta/T} - e^{2 \xi_B} \right),
\label{dot-nB-1}
\ee
where $d\tau_{l,\bar q} = d^3 p_l d^3 p_{\bar q} /[ 4 E_l E_{\bar q} (2\pi)^6]$. 
We assumed here that all participating particles are in kinetic equilibrium, i.e. their distribution functions have the 
form 
\be
f = 1/[\exp{(E/T - \xi}) + 1], 
\label{f-eq}
\ee
with $\xi = \mu/T$ being dimensionless chemical potential. 
In expression (\ref{dot-nB-1}) $\xi_B$ and $\xi_L$ denote baryonic and leptonic chemical potentials respectively and
the effects of quantum statistics are neglected but only for brevity of notations. Their effects are not essential in the sense that they do not change the 
conclusion. The assumption of kinetic equilibrium is well justified because it is enforced by the very efficient elastic scattering.
Another implicit assumption is the usual equilibrium relation between chemical potentials of particles and antiparticles,
$\bar \mu = -\mu$, imposed e.g. by the fast annihilation of quark-antiquark or lepton-antilepton pairs into two and three photons.
Anyhow the assumption of chemical equilibrium is one of the cornerstones of the spontaneous baryogenesis.
  
The conservation of $(B+L)$ implies the following relation: $\xi_L = - \xi_B/3$. Keeping this in mind, we find
\be
\dot n_B + 3H n_B \approx 
 -\left(1 - e^{ \dot \theta/T - 3\xi_B +\xi_L} \right) I  \approx   \left( \frac{\dot \theta }{T} - \frac{10}{3}\,\xi_B \right) I,
\label{dot-nB-2}
\ee
where we assumed that $\xi_B$ and $\dot \theta/T$ are small.
In relativistic  plasma with temperature $T$ the factor $I$, coming from the collision integral, 
can be estimated as $I=T^8/m^4$, where $m$ is a numerical constant with dimension of mass. 
It differs from $m_X$, introduced in eq.~(\ref{L-of-theta-1}), by a numerical coefficient.

The asymmetry between quarks and antiquarks having the distribution (\ref{f-eq}) with equal by magnitude but opposite by sign 
chemical potentials and identical dispersion relations is equal to
\be
n_B = C_B {\xi_B T^3} ,
\label{n-B-eq}
\ee
where $C_B$ is a constant,
see Eq.~(\ref{n-B-of-theta}) in the limit of $\mu \ll T$, because in the realistic case the baryon asymmetry is quite small.

For a large factor $I$ we expect the equilibrium solution $\xi_B = (3/10) \dot\theta/T$, so $\dot\theta$ up to the different numerical
factor seems to  be the baryonic chemical potential, as expected in the usually assumed SBG scenario. An emergence of the factor $3/10$
instead of $1/3$ in the equilibrium expression is due to the conservation law $B+L = const$. 
However, as we have seen above, the baryonic chemical potential is not aways  proportional to $\dot\theta (t)$.

\section{Kinetic equation for time-varying amplitude \label{s-kin-eq}}

In the case the interaction proceeds in a time dependent field and/or the time duration of the process is finite, then the energy conservation
delta-function in (\ref{dot-nB-1}) does not emerge and the described in Sec.~\ref{s-kin-eq-canon} approach becomes invalid,  so one has to make the time integration with an account of  time-varying background and integrate over the phase space without energy conservation.

In what follows we consider two-body inelastic process with baryonic number non-conservation with the amplitude obtained from the
last term in Lagrangian (\ref{L-of-theta-1}). At the moment we will not specify the concrete form of the reaction but only will say that it is
the two-body reaction 
\be
a+b \lrar c+d ,
\label{ab-go-cd}
\ee
where $a,b,c$, and $d$ are some quarks and leptons or their antiparticles. The expression for the 
evolution of the baryonic number density, $n_B$, follows from eq. (\ref{kin-eq-1}) after integration of its both sides over $d^3 p_i/(2\pi)^3$.
Thus we obtain:
\be
\dot n_B + 3 H n_B = -\frac{(2\pi)^3}{t_{max}}  \int d\nu_{in} d\nu_{fin}\,\delta({\bf P}_{in} -{\bf P}_{fin} )\, |A|^2 
\left( f_a f_b - f_c f_d \right)\,,
\label{nB-kin-eq}
\ee
where e.g. $d\nu_{in} = {d^3 p_a d^3 p_b}/{[ 4 E_a E_b (2\pi)^6}]$ and
the amplitude of the process is defined as
\be
A = \left(\int_0^{t_{max}} dt\, e^{i [ (E_c+E_d - E_a -E_b)t  + \theta (t) ]} \right) F(p_a,p_b,p_c,p_d)\, ,
\label{amp}
\ee
and $F$ is a function of 4-momenta of the participating particles, determined by the concrete form of the interaction Lagrangian.
In what follows we consider two possibilities: $ F= const$ and $F = \psi^4 \,m_X^{-2}$, where in the last case $\psi^4$ symbolically 
denotes the product of the Dirac spinors of particles $a, b, c$, and $d$.

In the case of equilibrium with respect to baryon conserving reactions the distribution functions have the canonical form,
$f_a = \exp (-E_a /T + \xi_a)$, where $\xi_a \equiv \mu_a/T$ is the dimensionless chemical potential. So for constant $F$
the product  $|A|^2 (f_a f_b - f_c f_d) $ depends upon the particle 4-momenta only through $E_{in}$ and $E_{fin}$, where
\be
E_{in} = E_a +E_b,\,\,\,\,{\rm and}\,\,\,\, E_{fin} = E_c+E_d .
\label{E-in-E-fin}
\ee  

To integrate Eq.~(\ref{nB-kin-eq}) over the phase space it is convenient to change the integration variables, according to:
\be 
\frac{d^3 p_a}{2 E_a}\,\frac{d^3 p_b}{ 2 E_b } =  d^4 P_{in}\, d^4 R_{in} \,\delta (P^2_{in} + R^2_{in} )\, \delta (P_{in}R_{in})\,, 
\label{pp-PR}
\ee
where $P_{in} = p_a +p_b$ and $R_{in} = p_a - p_b$ and masses of the particles are taken to be zero. Analogous expressions are
valid for the final state particles. Evidently the time components of the 4-vectors $P$ are the sum of  energies of the incoming and outgoing 
particles, $P^{(4)}_{in} = E_{in} $ and $P^{(4)}_{fin} = E_{fin} $. Now we can perform almost all (but one)  integrations and finally we obtain 
the kinetic equation in the following form:
\be
\dot n_B + 3 H n_B = - \frac{T^5}{2^5 \pi^6\, t_{max}}\, \int_0^\infty dy  \left[ e^{\xi_a + \xi_b} \left(  |A_+|^2 + |A_-|^2 e^{-y}   \right) -
e^{\xi_c+ \xi_d} \left(  |A_{ - }|^2 + |A_{+}|^2 e^{-y}   \right) 
\right] ,
\label{dot-nB-gen}
\ee
where $y=  E_-/T$ is the dimensionless energy with $E_-$ being the difference between initial and final energies of the system, $E_- = E_{in} - E_{fin} $,   
$A_+$ and $A_-$ are amplitudes taken at positive and negative $E_-$, respectively. Note, that with the substitution 
$E_- \rar |E_-|$  the only difference between $A_+$ and $A_-$ is that  $A_- (\theta) = A_+ (-\theta)$. 

The equilibrium is achieved when the integral in Eq.~(\ref{dot-nB-gen}) vanishes. Clearly it takes place at
\be
\xi_a+\xi_b -\xi_c -\xi_d = \frac{ \langle |A_+|^2 e^{-y} + |A_-|^2 \rangle } {  \langle |A_+|^2  + |A_-|^2 e^{-y} \rangle} -1 ,
\label{equiv-sol}
\ee
where the angular brackets mean integration over $dy$ as indicated in Eq.~(\ref{dot-nB-gen}).

This results above are obtained for the amplitude which does not depend upon participating particle momenta. 
The calculations would be be somewhat more complicated if this restriction is not true. For example if the baryon 
non-conservation takes place in  four-fermion interactions, then the amplitude squared can contain the terms of 
the form $(p_a p_b)^2 /m_X^4$ or  $(p_a p_c)^2 /m_X^4$, etc. The effect of such terms results in a 
change of the numerical coefficient in Eq.~(\ref{dot-nB-2}) but  the latter is unknown anyhow,
and what is more important the temperature coefficient in front of  the integral in this equation would change from $T^5$
to $T^9/m_X^4$.

\section{Examples of time-varying $\theta$ \label{s-examples}}

\subsection{Constant $\dot \theta$ \label{ss-const-dot-tjeta}}

This is the case usually considered in the literature and the simplest one. The integral (\ref{amp}) is taken analytically resulting in:
\be
|A|^2 \sim \frac{ 2- 2\cos[ ( \dot\theta - E_-) t_{max}]}{ (\dot\theta - E_-)^2}\, ,
\label{A-const-dot-theta}
\ee
where $E_-$ is running over the positive semi-axis.

For large $t_{max}$ this expression tends to $\delta (E_--\dot\theta)$, so $ |A_+|^2 = 2\pi \delta (E_-  - \dot\theta) t_{max} $ 
and $ |A_-|^2 = 2\pi \delta (E_-  +\dot\theta) t_{max} =0$, if $\dot\theta >0$ and vice versa otherwise.
Hence the equilibrium solution is
\be
\xi_a+\xi_b -\xi_c -\xi_d - \dot\theta = 0, 
\label{equil-1}
\ee
coinciding with the standard result.

The limit of $\dot\theta = const$ corresponds to the energy non-conservation by
the rise (or drop) of the energy of the final state in reaction (\ref{ab-go-cd}) exactly by $\dot\theta$.
However if  $t_{max}$  is not sufficiently large, the non-conservation of energy is not equal to $\dot\theta$ 
but somewhat spread out and the
equilibrium solution would be different. There is no simple analytical expression in this case, so 
 we have to take the integral (\ref{equiv-sol}) over $y$ numerically
to find at what values of chemical potentials, $\xi_k$, it vanishes and this point determines the equilibrium values of the chemical 
potentials in external $\dot\theta $ field.

\begin{figure}
\centering
 \includegraphics[width=.46\textwidth]{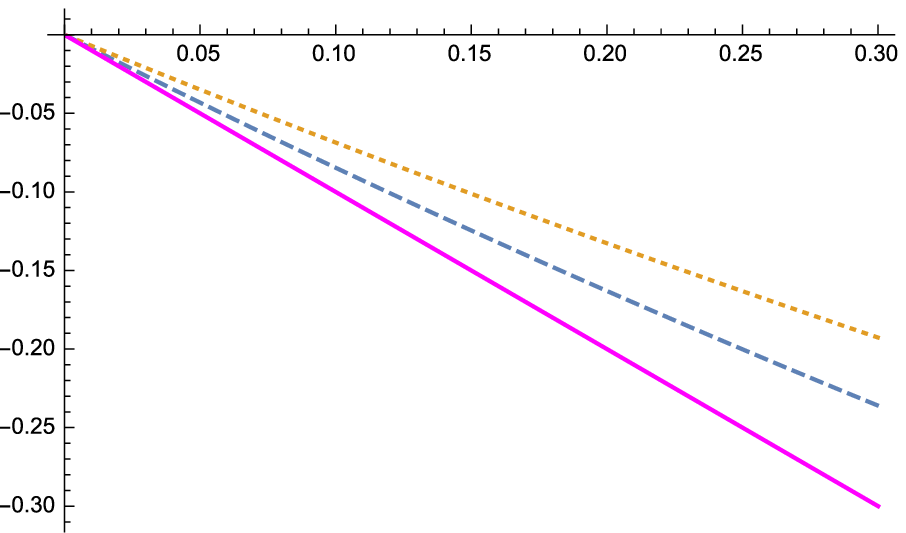} \hspace{.2cm}
\includegraphics[width=.46\textwidth]{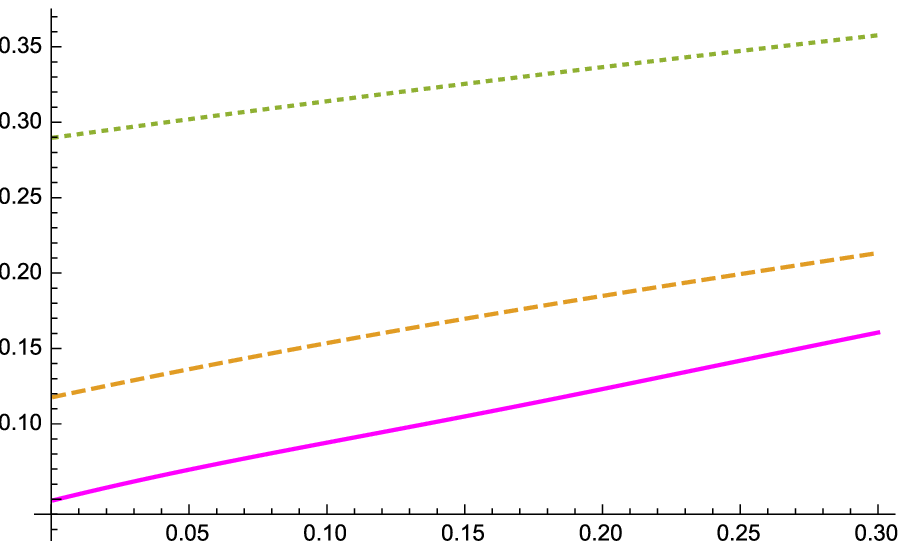}
\caption{
{\it Left}: $\dot \theta/T$ for infinite time integration (thick line). 
Cut-off in time integration $\tau \equiv t_{max} T =$ 10 (dashed);   3 (dotted).
{\it Right}: relative differences between  the  equilibrium solutions and $\dot\theta/T$, normalized to $\dot\theta/T$,
as functions of $\dot\theta$ for different $t_{max}$:  
 $\tau \equiv t_{max} T =$ 30 (thick);   10 (dashed);  3 (dotted).}
\label{fig-1}       
\end{figure}

The results of the calculations are presented in Fig.~\ref{fig-1}. In the left panel the values of the r.h.s. of Eq.~(\ref{equiv-sol})
are compared with $\dot\theta/T$ (thick line) for two values of the cut-off in time integration $\tau \equiv t_{max} T = 10$ (dashed line) and $\tau = 3$ (dotted line).   
In  the right panel relative differences between  the  r.h.s. of Eq.~(\ref{equiv-sol}) and $\dot\theta/T$, normalized to $\dot\theta/T$,
as functions of $\dot\theta$ for different
maximum time of the integration are depicted. 
We see that for $\tau = 30$ (thick line) the deviations are less than 10\%,
while for $\tau = 3$ (dotted line) the deviations are about 30\%. If we take $\tau$ close to unity, the deviations are about 100\%. 
The value of $\dot\theta/T$ is bounded from above by 0.3 because at large $\dot\theta/T$ the linear expansion, used
in our estimates, is invalid.

\subsection{Second order Taylor expansion of $\theta (t)$ \label{ss-Taylor}}

Here we assume that $\theta (t)$ can be approximated as 
\be
\theta (t) = \dot \theta \,t + \ddot \theta\, t^2/2,
\label{theta-Taylor}
\ee
where $\dot\theta$ and $\ddot\theta$ are supposed to be constant or slowly varying. 
In this case the integral over time (\ref{amp}) can also be taken analytically but the result is rather complicated.
We need to take the integral
\be
\int_0^{t_{max}} dt \exp [i \theta (t) ] .
\label{int-theta-of-t}
\ee
Its real and imaginary parts are easily expressed through the Fresnel functions.  So the amplitude squared is given by the
functions tabulated in Mathematica and the position  of the equilibrium point can be calculated, as in the previous case, 
by numerical calculation of one dimensional integral.

\begin{figure}
\centering
 \includegraphics[width=.46\textwidth]{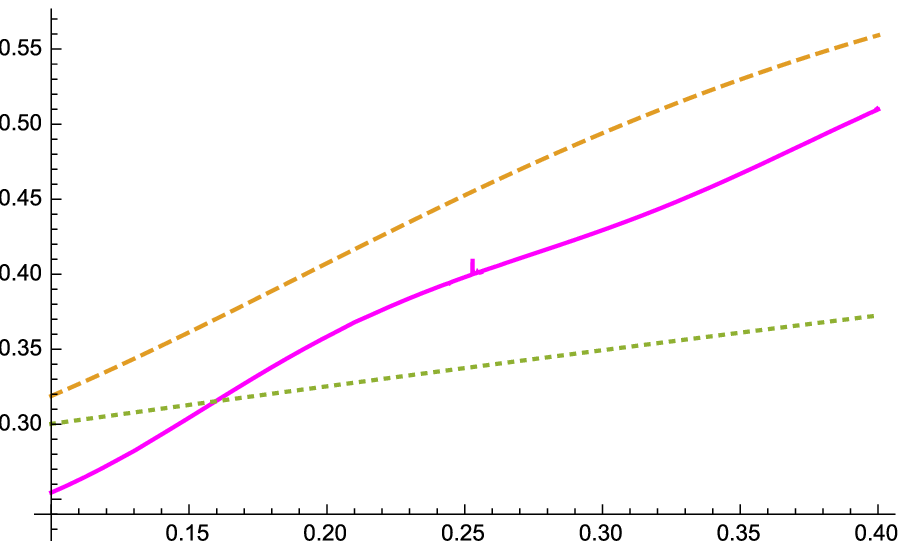} \hspace{.2cm}
\includegraphics[width=.46\textwidth]{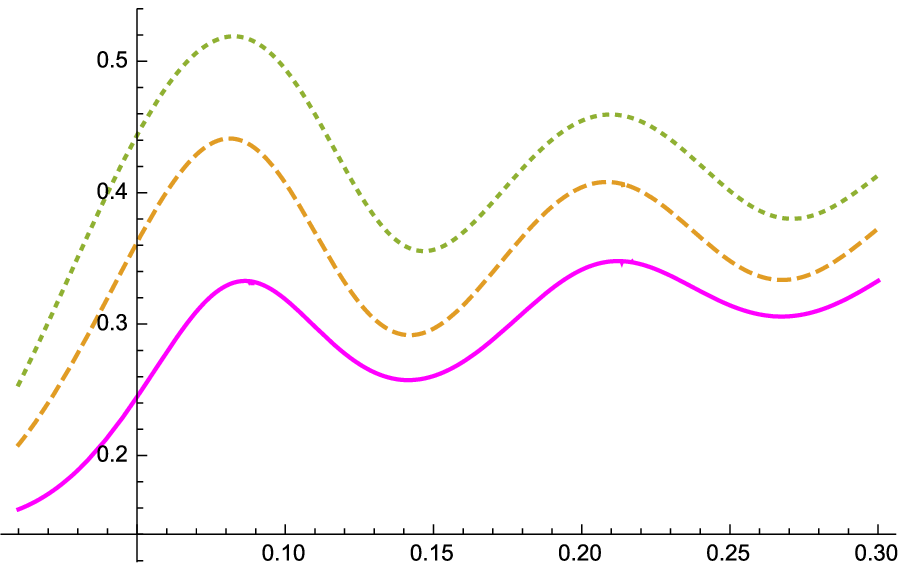}
\caption{
Relative differences between   equilibrium solutions and $\dot\theta/T$, normalized to $\dot\theta/T$. 
{\it Left:}
as functions of $\dot\theta$ for different $t_{max}$: 
  $\tau \equiv t_{max} T = 30$ (thick);  10 (dashed);  3 (dotted); $\ddot \theta = 0.1$.
 {\it Right:} 
as functions of $\ddot\theta$ for different $\dot \theta$:  
 $\dot \theta =$  0.1 (thick); 0.2 (dashed);   0.3 (dotted); $\tau \equiv t_{max} T= 10$.}
\label{fig-2}      
\end{figure}

The r.h.s. of Eq.~(\ref{equiv-sol}) as functions of $\dot\theta $ for different values of $\tau$ are presented in Fig.~\ref{fig-2}, left panel. It is 
interesting that the dependence on $\tau$ is not monotonic. This can be explained by that at small $\tau$ the effects of
$\ddot\theta t^2$ are not essential. 

To check the dependence on $\ddot\theta$ we calculated again the  r.h.s. of Eq.~(\ref{equiv-sol}) but now as functions of 
$\ddot\theta$ presented in the right panel in Fig.~\ref{fig-2}. We see that the equilibrium point oscillates as a function of $\ddot\theta$. 

\section{Conclusion \label{s-conclude}}

We argue that in the standard
description $\dot\theta$ is not formally the chemical potential, though in thermal equilibrium 
$\mu_B$ tends to $\dot \theta$ with numerical, model dependent,
coefficient.
Moreover,  this is not always true but depends upon the chosen representation for the "quark"
fields. In the theory described by the Lagrangian (\ref{L-of-theta-1}) which appears "immediately" after the spontaneous symmetry 
breaking,  $\theta (t)$ directly enters the interaction term in this Lagrangian and in equilibrium  $\mu_B \sim \dot\theta$ indeed. On
the other hand, if we transform the quark field, so that the dependence on $\theta$ is shifted to the bilinear product of the quark 
fields (\ref{L-of-theta-2}), then chemical potential in equilibrium does not tend to $\dot\theta$, but  to zero. On the other hand, the 
magnitude of the baryon asymmetry in equilibrium is always proportional to $\dot\theta$. It can be seen, according to the equation
of motion of the Goldstone field, that $\dot\theta/T$ drops down
in the course of the cosmological cooling as $T^2$, so the baryon number density in the comoving volume decreases in the
same way. So to avoid complete vanishing of $n_B$ the baryo-violating interaction should switch-off at some non-zero $T$. This is
always the case but the dependence on the interaction strength is non-monotonic. 

The assumption of  constant or slowly changing $\dot\theta$, which is usually done in the SBG scenario, may be not fulfilled and
to include the effects of an arbitrary variation of $\theta (t)$ as well as the effects of the finite time integration we transform the kinetic 
equation in such a way that it becomes operative in the case of non-conserved energy. A shift of the equilibrium value of the baryonic
chemical potential due to this effect is numerically calculated. 

\vspace{.5cm}
{ \bf Acknowledgement.}
EA and AD thank the support of the Grant of President of Russian Federation for the leading scietific schools 
of the Russian Federation, NSh-9022.2016.2. VN thanks the support of the Grant RFBR 16-02-00342.


\begin{thebibliography}{}

\bibitem{spont-BG-1}
A. Cohen, D. Kaplan, Phys. Lett. \textbf{B 199}, 251 (1987).

\bibitem{spont-BG-2}
A. Cohen, D. Kaplan, Nucl.Phys. \textbf{B308}, 913 (1988).

\bibitem{spont-BG-3}
A. G. Cohen, D.B., A.E. Nelson, Phys.Lett. \textbf{B263}, 86-92 (1991).

\bibitem{BG-rev}
A.D.Dolgov, Phys. Repts \textbf{222} (1992) No. 6;\\
V.A. Rubakov, M.E. Shaposhnikov, Usp. Fiz. Nauk \textbf{166}, 493 (1996); \\
A. Riotto, M. Trodden, Ann. Rev. Nucl. Part. Sci. \textbf{49}, 35 (1999); \\ 
M.~Dine, A.~Kusenko, {Rev. Mod. Phys.} \textbf{76}, 1 (2004).

\bibitem{AD-30}
A.D. Dolgov, Surveys in High Energy Physics \textbf{13}, 83 (1998). 

\bibitem{ad-zeld-cpt}
A.D. Dolgov, Ya.B. Zeldovich, Uspekhi Fizicheskih Nauk \textbf{130}, 559 (1980); Rev. Mod. Phys. \textbf{53}, 1-41 (1981).

\bibitem{ad-cpt}
A.D. Dolgov, Phys. Atom. Nucl. \textbf{73}, 588-592 (2010). 


\end{thebibliography}
\end{document}